ATA Memo #51

# Customized Beam Forming at the Allen Telescope Array

G. R. Harp
August 11, 2002

## Abstract

One of the exciting prospects for large N arrays is the potential for custom beam forming when operating in phased array mode. Pattern nulls may be generated by properly weighting the signals from all antennas with only minor degradation of gain in the main beam. Here we explore the limits of beam shape manipulation using the parameters of the Allen Telescope Array. To generate antenna weights, we apply an iterative method that is particularly easy to understand yet is comparable to linearly-constrained methods. In particular, this method elucidates how narrow band nulls may be extended to wider bandwidth. In practical RFI mitigation, the gain in the synthetic beam is obviously affected by the number and bandwidth of nulls placed elsewhere. Here we show how to predict the impact of a set of nulls in terms of the area of sky covered and null bandwidth. Most critical for design of the ATA, we find that high-speed (~10 ms) amplitude control of each array element over the full range 0-1 is critically important to allow testing of wide area / wide bandwidth nulling.

## 1. Introduction

Man-made radio frequency interference (RFI) is becoming an increasingly difficult problem for radio astronomy. In their frequency bands, satellite radio transmissions or ground-based radar and wireless communications signals can easily swamp extraterrestrial sources. Historically, RFI has been ameliorated by limiting astronomical observations to bands where no man-made sources exist (frequency blanking) and / or by performing astronomical observations when the man-made sources are turned off (time blanking). As the number of RFI sources grows, the time / frequency parameter space available for astronomical observations decreases rapidly, which has led to a search for new solutions.

This paper focuses on RFI removal in the beamformer of a large N array (i.e. in phased array mode). Apart from time / frequency blanking, in such systems RFI mitigation techniques might be classified into two additional categories[1]: 1) adaptive schemes which use an estimate of the RFI source signal and remove it from the observation signal, and 2) null steering methods which minimize the gain of the array in the direction of the interferer (also known as direction of arrival methods). The adaptive schemes are typically applied *after* the beamforming process, whereas null steering manipulates the signals before they enter the beamformer. All these strategies show real promise, and will play a role in the future RFI mitigation strategies of general-purpose radio astronomy observatories.

Here we discuss a novel approach to the null steering method. Previous treatments (for an excellent review, see Ref. 1) employ linearly-constrained (i.e. matrix) methods to place point nulls in the directions of interferers. In theory these nulls are perfect (more than 100 dB suppression can be achieved if desired, see e.g. Ref. 4) and in practice the null depths are limited only by amplitude and phase calibration of the antenna data streams.

Here we describe a heuristic null steering approach. In application, this approach gives results comparable to linearly-constrained methods and calculations can be performed in real time. However, the greatest advantage of this approach is pedagogical, as it is quite easy to see how to generalize point nulls to wide area and / or wide bandwidth.

From studies with the iterative approach we show that when point nulls are grouped together, the total nulled area is larger than might be naively expected. Generalizing from narrow to wide bandwidth is straightforward, and it is shown that the linear dimension of a narrow band null can be traded off against frequency bandwidth (BW). We propose a single metric describes the total volume of parameter space that is occupied by any null region.

It is well-known that the shape of the central beam region can be manipulated for specific applications. We consider the potentially useful shape of a “ring beam”, consisting of a narrow ring (0.05° at 1.4 GHz) and a deep central null. Coupled with an ordinary beam, the ring beam may useful for simultaneous on / off studies of point sources, as in SETI searches or pulsar studies. We give an example of such a shaped beam to demonstrate the flexibility obtainable within the parameters of the ATA.

Most importantly, this paper explores the limits of custom beam forming that can be achieved with the baseline design of the Allen Telescope Array (ATA). We assume the ATA is composed of $N = 350$ identical dishes, arrayed over an area approximately 1 km in diameter. We shall use a set of array positions designated “L8a_2” and generated by D. C.-J. Bock[2], which has been optimized to reduce near-in sidelobes within the site constraints at the Hat Creek Radio Observatory. Bock’s array does not specify the vertical displacements of antennas, so each antenna was assigned a $z$ coordinate according to a gaussian distribution with standard deviation of 2 m. The ATA frequency range of this array is ~ 0.5-11.2 GHz. See Ref. 3 for a thorough description of the baseline design.

Section 2 of the present paper describes an approximate implementation for the iterative approach which is particularly easy to understand. Section 3 develops an expression for the beamformer gain. Section 4 introduces the exact iterative method for placing a single point null, which is extended in section 5 to multiple nulls at a narrow BW. Section 6 describes wide BW (WBW) nulls. Section 7 considers the effects of gain or phase miscalibration and section 8 considers the possibility of shaping the regions of high gain with the ring beam. Sections 9 and 10 comprise the discussion and conclusion. Detailed examples are provided throughout the paper.

## 2. Iterative Approach

To motivate the mathematical development, we first present a visualization of the method in Figure 1. In this Section, a few approximations are made: the antenna positions are assumed coplanar and they are displaced from their exact positions by a small distance (<1 m) to force them onto a rectangular 512 × 512 grid. Here, as in all examples we assume the antennas have an omnidirectional response.

For simplicity, we assume the array is pointed at zenith (about 40° declination) and delays are applied to each antenna data stream to bring them into phase at zenith. Under these conditions, a simple fourier transform of the array positions (step 1 in the figure) with unity weighting provides an image of the naturally weighted beam (step 2).

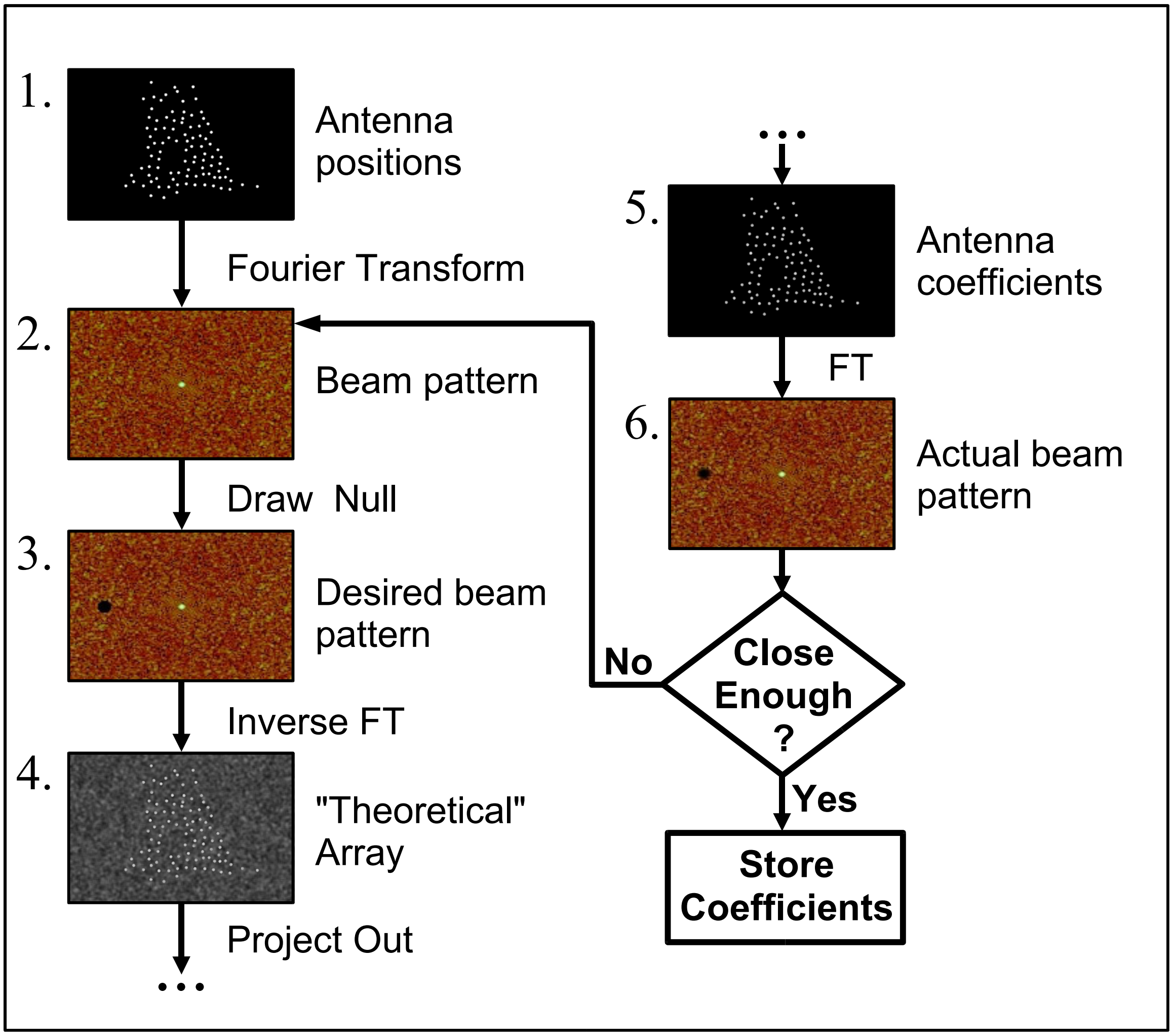

**Figure 1: Outline of an approximate iterative method for generating wide area nulls.**

We seek a set of coefficients that, when applied to the $N$ data streams coming from each antenna, will suppress the array gain in the nulling direction while preserving high gain in the direction of the original beam maximum. To do this, we "draw" our desired null region onto the beam pattern; that is, we replace with zeros those pixels where we wish to reduce gain. This results in step 3 of the figure, the desired beam pattern.

An inverse fourier transform gives the exact array coefficients necessary to create this pattern (step 4). However, this "theoretical array" contains complex amplitude over the entire antenna plane, particularly in regions where no antennas exist. So we approximate this theoretical array by projecting out the values where real antennas exist (step 5).

The values in step 5 are the first approximation to the antenna coefficients we need. A fourier transform of these coefficients shows the actual beam pattern in step 6. After just one iteration, the actual beam pattern may be a poor approximation to the desired beam pattern. If so, then the pattern of step 6 is fed back into the place of the original beam pattern in step 2 and the process is repeated.

Provided the null region is not too large (more on this below), our experience shows that each iteration will reduce the gain in the region of the null. Iterations continue until a desired null depth is reached (or in case of overly large null regions, the calculation blows up).[*]

## Example

Figure 2 shows a null that was calculated in exactly this way. This calculation was performed at 1420 MHz. At figure center there is the strong peak of the synthetic beam (1.2' in diameter). The white circle surrounding this peak shows the FWHM of the ATA primary beam calculated from[3] $3.5 / f$ (GHz). The null, which is about 20' in diameter, is clearly visible on the left hand side of the figure. The color scale is logarithmic, and the color bar on the right spans 50 dB in gain (units of power) where white (0 dB) corresponds to the beam maximum and black is –50 dB or less.

Figure 3 shows line scans through the original beam and that of Figure 2. These line scans were sampled at 4x the pixel spacing to fully map the beam pattern, and are normalized to unity gain at beam center. Here the region of the null is suppressed by ~30 dB relative to the original beam, and is > 50 dB below the main peak. This suppression comes at the price of a 20% broadening of the central peak ($FWHM_{orig} = 0.022°$, $FWHM_{null} = 0.027°$) and a 60% loss of signal to noise ratio. But notice that the area of the nulled region was chosen to be quite large to improve visual effect. In a real application the null would be smaller, perhaps focused on a single point. We postpone detailed analysis until after introduction of the exact formulation, below.

[*] It should be pointed out that we have not proven this approach is absolutely convergent, but subject to the constraints mentioned below, it has always been well-behaved.

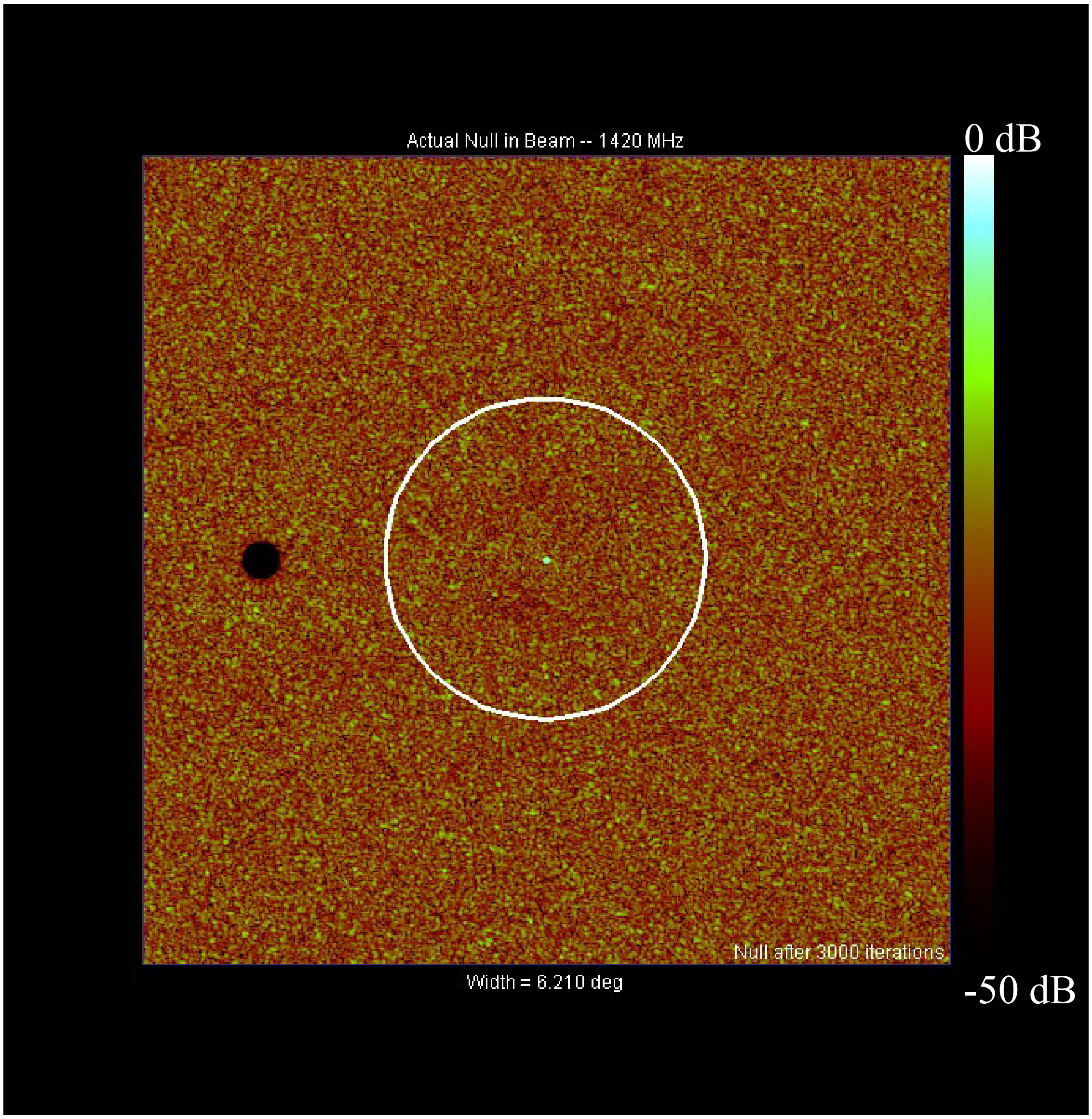

**Figure 2: The beam pattern with a 20' diameter null inserted to the left of beam center. This pattern was created using the parameters of the ATA and the approximate iterative method. The color scale is logarithmic as indicated.**

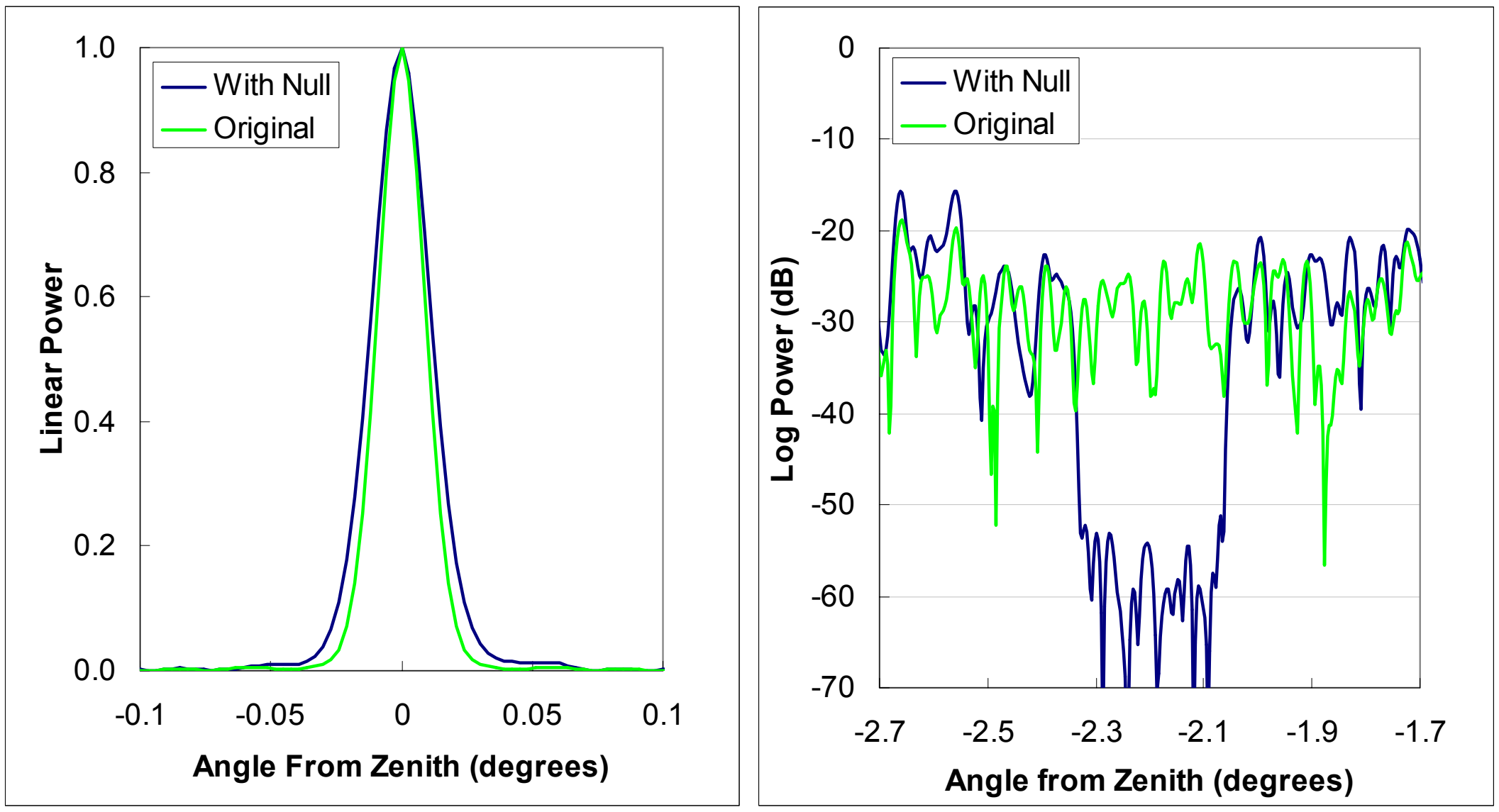

**Figure 3: Line scans through the main peak (left) and nulled region (right) of Figure 2. In the right-hand panel, 0 dB represents the gain at beam center. The calculations were performed at 1420 MHz.**

# 3. Beamformer Equation

We begin with a development of the fundamental equation for the output of a beamformer connected to $N$ identical antennas. The voltage response at frequency $f$ of a single antenna placed at the origin to a single point source may be written

$$dV(k,\hat{k},\hat{k}_0,t) = A(\vec{k}-\vec{k}_0)\, E(\vec{k},t)\, d\hat{k} \tag{1}$$

where $A(\vec{k}-\vec{k}_0)$ is the antenna gain (assumed same for all antennas), $\hat{k}$ is the direction to the source point, $\hat{k}_0$ is the antenna pointing direction, and $k = \dfrac{2\pi f}{c}$. The wavevectors $\vec{k}$ and $\vec{k}_0$ are defined as $\vec{k} = (k_x, k_y, k_z) \equiv k\hat{k}$ and $\vec{k}_0 = (k_{x0}, k_{y0}, k_{z0}) \equiv k\hat{k}_0$ (for now, note that both $\vec{k}$ and $\vec{k}_0$ refer to the same frequency). Finally, $E(\vec{k},t)$ is the electric field arriving at the antenna from direction $\hat{k}$ (we consider only a single polarization so $E$ is a scalar).

Integrating over all directions on the sky, the measured voltage from this antenna over an infinitesimal BW is

$$V(\vec{k}_0,t) = \int_{4\pi} A(\vec{k}-\vec{k}_0)\, E(\vec{k},t)\, d\hat{k}\,. \tag{2}$$

For another antenna at position $\vec{r}_i$

$$V_i(\vec{k}_0,t) = \int_{4\pi} A(\vec{k}-\vec{k}_0)\, E\left(\vec{k}, t-\tau_{gi}(\hat{k})\right) d\hat{k}\,, \tag{3}$$

where $\tau_{gi}(\hat{k}) = \frac{\hat{k}\cdot\vec{r}_i}{c}$ is the geometric delay for direction $\hat{k}$.

Before the voltages can be summed in the beamformer, they must be phased up in the direction of interest, usually in the direction the antennas are pointing. This is accomplished by applying an instrumental delay $\tau_i(\hat{k}_0) = \frac{\hat{k}_0\cdot\vec{r}_i}{c}$ to the voltage arriving from each antenna

$$V_i(\vec{k}_0,t) \Rightarrow \int_{4\pi} A(\vec{k}-\vec{k}_0)\, E\left(\vec{k}, t-\tau_{gi}(\hat{k})+\tau_i(\hat{k}_0)\right) d\hat{k}\,. \tag{4}$$

Notice that we have put the instrumental delay *ahead* rather than behind the down conversion stage. If the delay were applied after down conversion, it would be necessary also to correct for the local oscillator phase and the fringe phase caused by the LO traversing delay $\tau_i$. Here as throughout the paper, we ignore the effects of down conversion and treat the beamformer as if it were operating at the sky frequency.

Because of the narrow BW, the electric field may be written

$$E(\vec{k},t) = E(\vec{k})\, e^{i\omega t} \tag{5}$$

where $\omega = 2\pi f$. With this substitution the measured voltage becomes

$$V_i(\vec{k}_0,t) = e^{i\omega t} \int_{4\pi} A(\vec{k}-\vec{k}_0)\, E(\vec{k})\, e^{-i\left(\vec{k}-\vec{k}_0\right)\vec{r}_i}\, d\hat{k}\,. \tag{6}$$

At wavenumber $k$, the beamformer output $b(k,t)$ is a simple sum of all the $V_i$ with complex weighting factors $\omega_i(k,t)$:

$$b(k,t) = e^{i\omega t} \int_{4\pi} A(\vec{k}-\vec{k}_0)\, E(\vec{k}) \left[\sum_i \omega_i(k,t)\, e^{-i\left(\vec{k}-\vec{k}_0\right)\vec{r}_i}\right] d\hat{k}\,. \tag{7}$$

A real beamformer is sensitive over a finite BW, so the actual beamformer output is

$$B(t) = \int H(k)\, b(k,t)\, dk\,, \tag{8}$$

where $H(k)$ is the frequency response of the signal processing electronics. This is the beamformer equation.

The coefficients $\omega_i(k,t)$ vary slowly as a function of time (characteristic time scale ~100 ms at the ATA[4]), and we restrict our discussion to time periods over which they can be considered constant. Such times still represent millions of cycles of the radio wave. In addition, most beamformer designs allow application of only a single $\omega_i(k)$ over the entire BW measured by the beamformer, so here we assume the $\omega_i$ are merely numbers.

### Beam Gain

In Equation 7, the term in brackets represents the sensitivity (i.e. gain) of the beamformer to signals arriving with wavevector $\vec{k}$ and we denote it $S(\vec{k}-\vec{k}_0)$:

$$S(\vec{k}-\vec{k}_0) = \sum_i \omega_i \, e^{-i(\vec{k}-\vec{k}_0)\vec{r}_i} \quad . \tag{9}$$

To get the largest signal to noise ratio in the direction $\hat{k} = \hat{k}_0$ with no imposed nulls, all the weighting factors are unity. In this case, the peak $S(\vec{k}-\vec{k}_0)$ has a maximum of $N$, independent of $k$ (frequency).

## 4. Single Point Null

We seek coefficients, $\omega_i$, such that our beamformer will have low gain in the direction $\hat{k}_1$ while maintaining a high gain in the direction $\hat{k}_0$. For a single point null, $\omega_i$ is hardly changed from $\omega_i^0$, the coefficients for the beam with no null, so we write

$$\omega_i = \omega_i^0 - \delta_i^0 \, . \tag{10}$$

As the problem is stated, the solution for $\delta_i$ is not unique. One especially useful solution (obtained by inspection) is

$$\delta_i^0 = \frac{S^0(\vec{k}_1-\vec{k}_0)}{N} \, e^{i(\vec{k}'-\vec{k}_0)\vec{r}_i} \quad , \tag{11}$$

where $S^0(\vec{k}_1-\vec{k}_0)$ is the original uniformly weighted gain in the direction we wish to null. This solution gives zero gain in the direction $\hat{k}_1$.

### Example

Figure 4 shows the result of a calculation of the beam gain before and after placement of a point null 50° away from the beam peak. The calculation was performed at 1420 MHz and the beam peak was at zenith. On the left we show a line scan through the null, whose full width at –50 dB is 14”. This is quite similar to the results of Ref. 4.

A calculation of the beam pattern as a function of $k$ was performed, and is plotted in frequency units on the right hand side of the figure. The width of the null at –50 dB is only ~ 50 kHz. Again, this is similar to the results of Ref. 4.

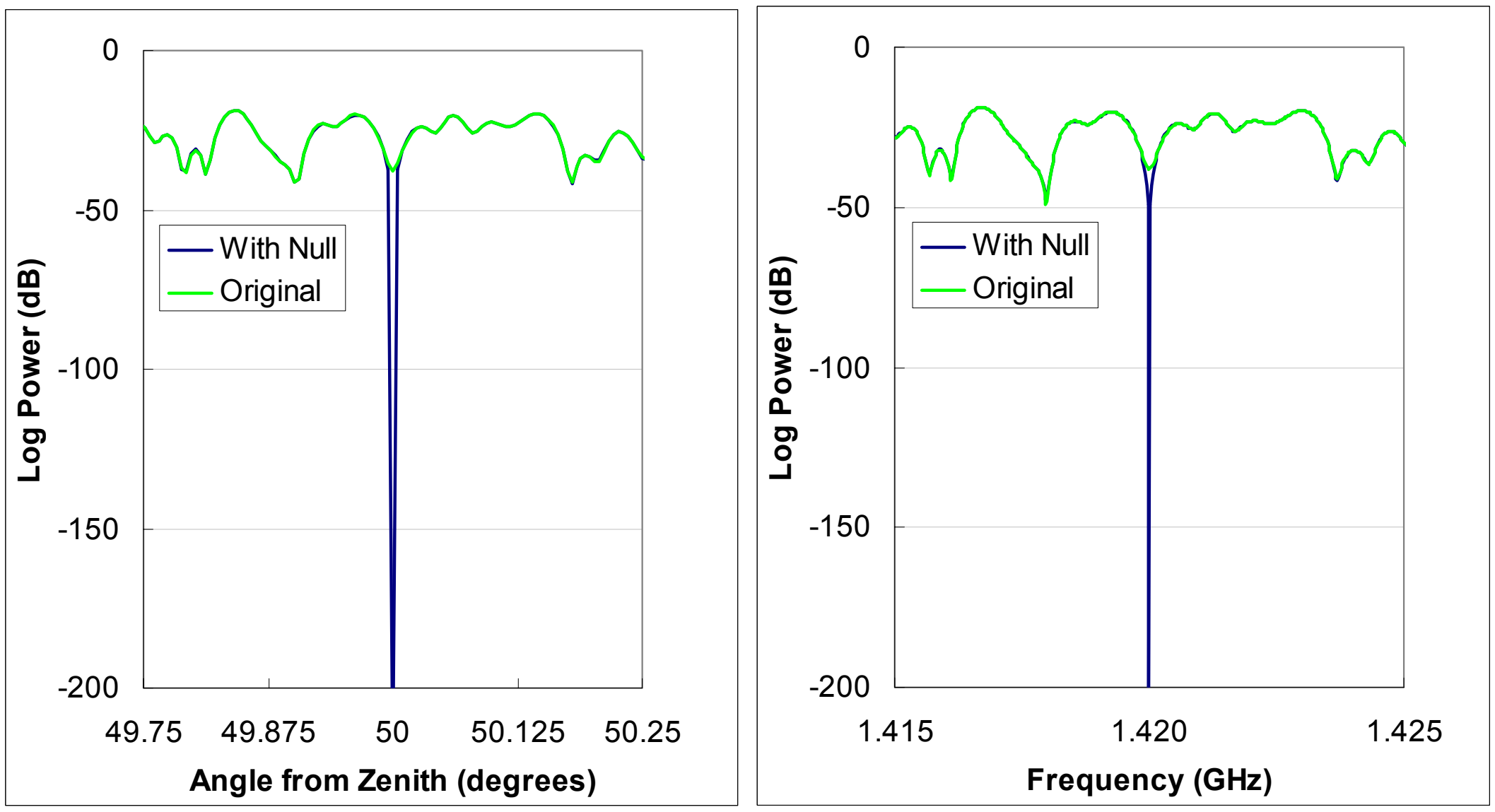

**Figure 4: Line scan as a function of zenith angle (left) and frequency (right) of the single point null described in the text. Notice the remarkable similarity of the scan shapes in the two panels.**

The scans in the two panels of Figure 4 look remarkably similar, considering they are plotting gain versus different variables. To see the physics behind this, we choose our coordinate system to put the null in the $xz$ plane. Then the left panel can be regarded as showing (approximately) $S(k_x)$, whereas the right panel shows $S(k)$. Now note that if all the antennas lie in a plane perpendicular to the $z$ axis, then $S$ is independent of $k_z$, resulting in $S(k_x) \approx S(k)$. In the present case the variance of the $z$ coordinate is 2 m, so we expect this similarity to break down when $\Delta k_z\,(2\text{m}) \sim 1$ radian, or $\Delta f \sim 40$ MHz.

# 5. Multiple Nulls

The extension to $M$ nulls is straightforward. We anticipate an iterative solution and define the beam gain $S^p(\vec{k}, \vec{k}_0)$ after $p$ iterations recursively:

$$S^p(\vec{k} - \vec{k}_0) = \sum_i \omega_i^p \; e^{-i\left(\vec{k} - \vec{k}_0\right)\vec{r}_i} \quad , \tag{12}$$

where $\omega_i^p$ is the antenna coefficient for antenna $i$ after $p$ iterations and

$$\omega_i^p = \omega_i^{p-1} - \frac{1}{N} \sum_{m=1}^{M} S^{p-1}(\vec{k}_m - \vec{k}_0) \; e^{i\left(\vec{k}_m - \vec{k}_0\right)\vec{r}_i} \; . \tag{13}$$

The sum over $m$ is performed over all null points. As a reminder, note that $|\vec{k}_m| = |\vec{k}_0| = k$.

In the case of $M > 1$, the nulls are not perfect after one iteration. At the $m^{\text{th}}$ null, all of the original amplitude is removed in one iteration, but the corrections made for other nulls introduce new gain into direction $\hat{k}_m$. Provided the number of nulls is not too large, each iteration tends to decrease the gain in the region of the null. We justify this statement with two empirical studies.

## Example: Randomly Placed Nulls

As more and more nulls are placed, the gain at the pattern center degrades. This degradation can be quantified with the signal to noise ratio (SNR) calculated from

$$\mathrm{SNR} = \frac{\left|\sum_i \omega_i\right|^2}{\sum_i |\omega_i|^2}. \qquad (14)$$

Here we are assuming that the synthesized beam is pointed directly at an unresolved point source whose signal we wish to measure. For a uniformly weighted beam, the SNR is $N$ times larger than for a single antenna, as expected. In Figure 5 this SNR is plotted versus the number nulls. These nulls were placed randomly at isolated positions over the sky hemisphere.

The SNR decreases linearly as the number of nulls is increased. We note that if more than 340 null points are placed, then the iterative method does not converge. Such a result can be understood from the fact that there are only 349 antenna weights available for adjustment, or 349 degrees of freedom. Each null consumes 1 degree of freedom. We compare these results to those of Bower (Ref. 4) with the curve labeled "Linearly-constrained". Figure 5 validates the present technique as giving results comparable to other methods.

When the nulls are placed randomly, they generally have a diameter and BW similar to that in Figure 4. A 14" diameter null has area (that is, solid angle) of 0.04 sq. arc min. From this and Figure 5, we can estimate the maximal area of the sky which can be nulled. Choosing SNR = 250 as a bounding limit, we estimate that no more than 4 square arc minutes can be nulled at a time (diameter ~ 2'), at 1420 MHz. In the next section, we shall see that this estimate is badly wrong.

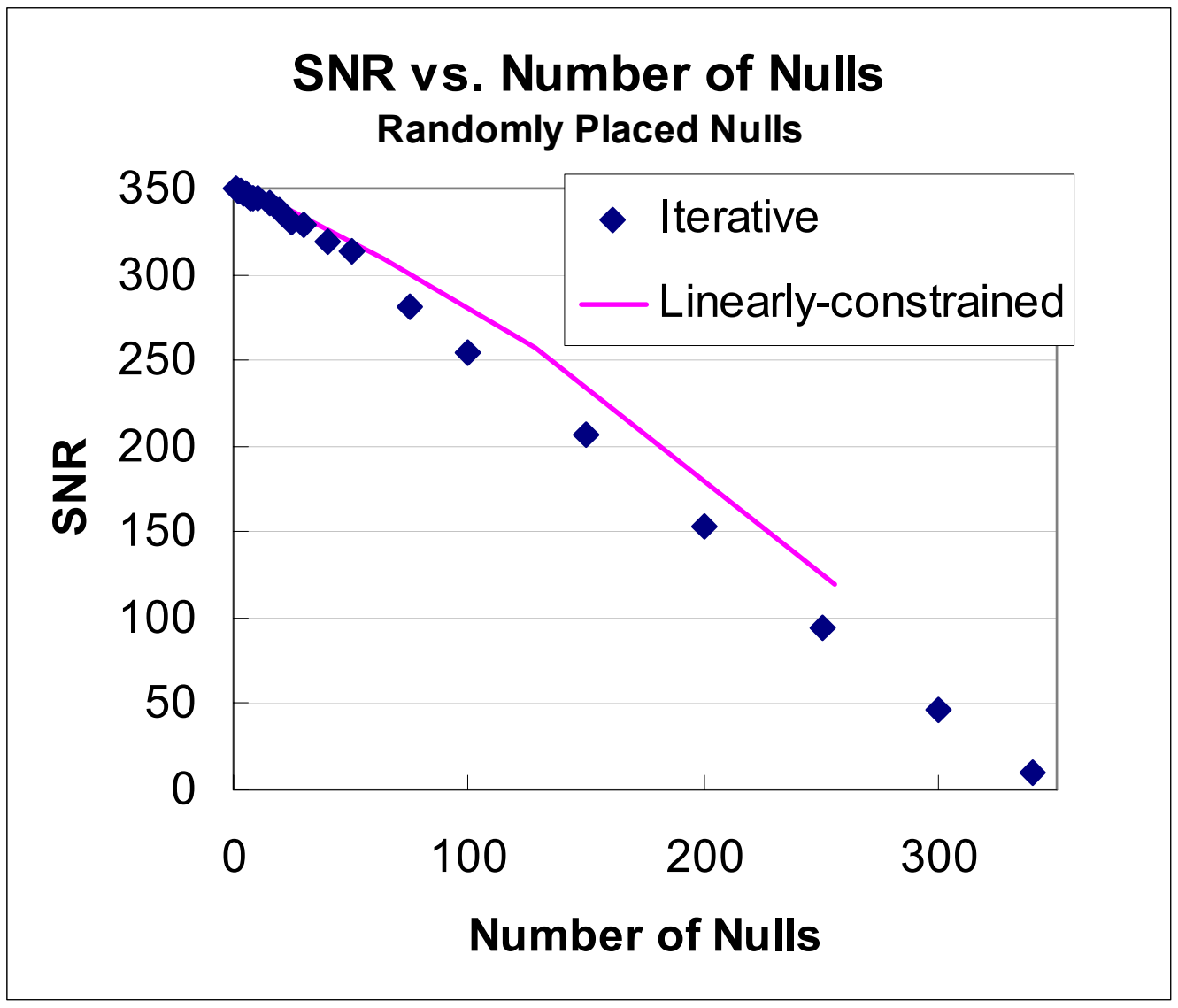

**Figure 5: Signal to noise ratio of the ATA when the synthesized beam is placed on an unresolved point source, as a function of the number of nulls placed on the sky. These nulls are placed at random, isolated points. The present, iterative method is compared with similarly prepared results of a linearly-constrained approach.**

## Example: Wide Area Nulls

In Figure 6 we present a beam pattern having a central peak is at zenith and a null in the $xz$ plane at zenith angle 50°. The null was specified to be 6' in diameter (28 sq. arc min.) though its actual full width at –50 dB is 9'.

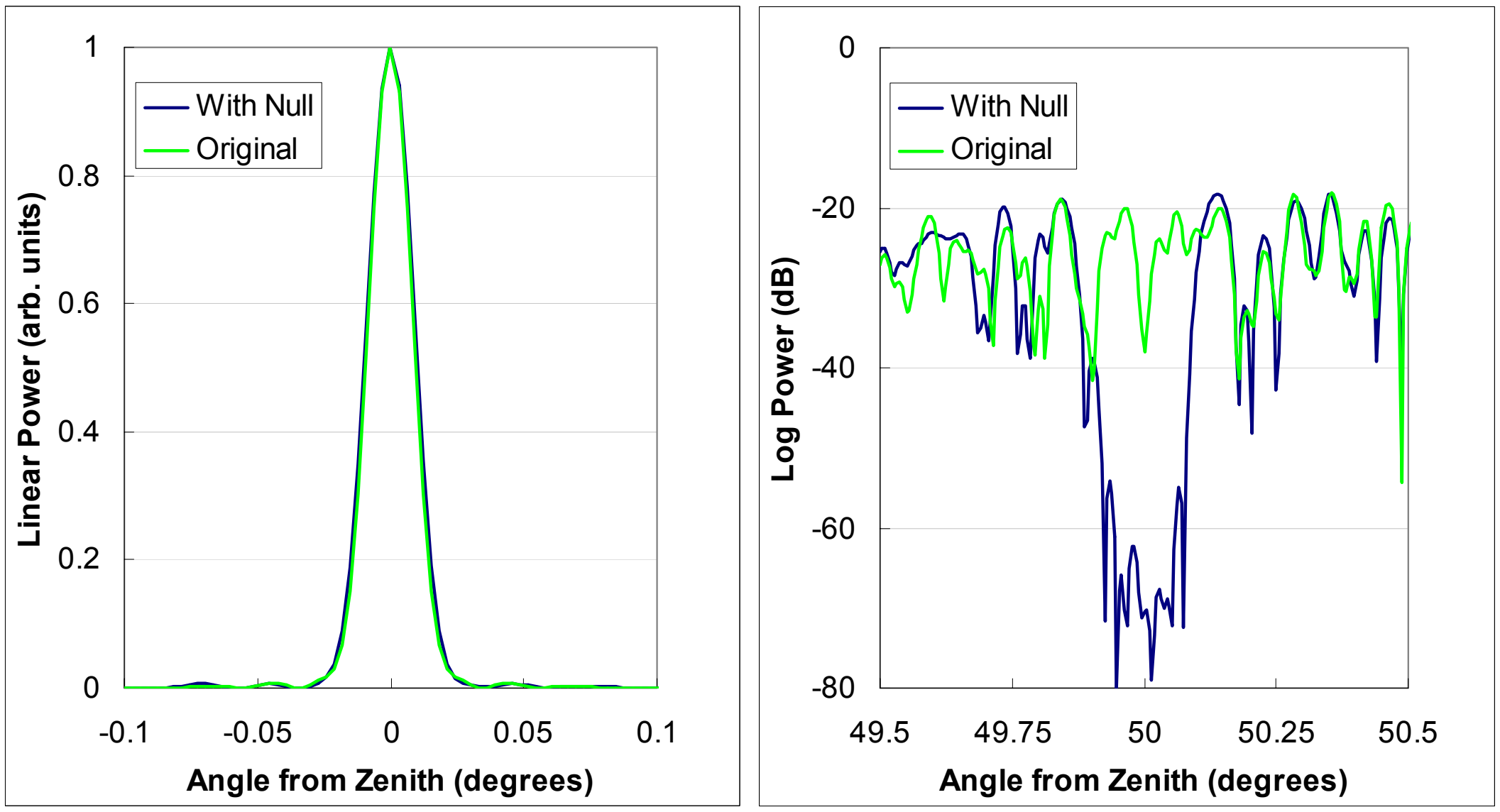

**Figure 6: Line scans through the central peak (left) and nulled region (right) for a 6' diameter null placed at 50° zenith angle.**

The curves in Figure 5 show the gain pattern after 251 iterations. In this case, the FWHM of the main peak is hardly changed, and has an SNR = 305 (87% of uniformly weighted array). At this SNR level, the present null has a much larger area than thought possible based on the previous section. It was accomplished by placing a grid of only 54 null points over the circular region. The spacing between nulls is 0.7'; we find empirically that this spacing is close enough to ensure that the null amplitude does not rise significantly between null points. In a loose way, we can describe these closely spaced nulls as "occupying" an effective area of about 2 sq. arc min. What we mean by this is that to reduce the pattern gain over a large area to less than –50 dB, we can use the effective area to determine the packing density.

One can come to an estimate of the effective area for a single null by analogy with the Nyquist theorm. We intuitively guess that when nulling a wide area, it is necessary to place point nulls with a separation no larger than

$$\Delta k = \frac{2\pi}{L}, \qquad (15)$$

where $L$ is the array diameter. Setting $L$ = 1000 m[†] leads to a point spacing of 0.73' (indeed, this influenced our choice of point spacing above). We find that when creating wide area nulls, Eq. 15 always provides a good estimate for the appropriate point spacing. We shall expand on the topic of point spacing in sections 6 and 9.

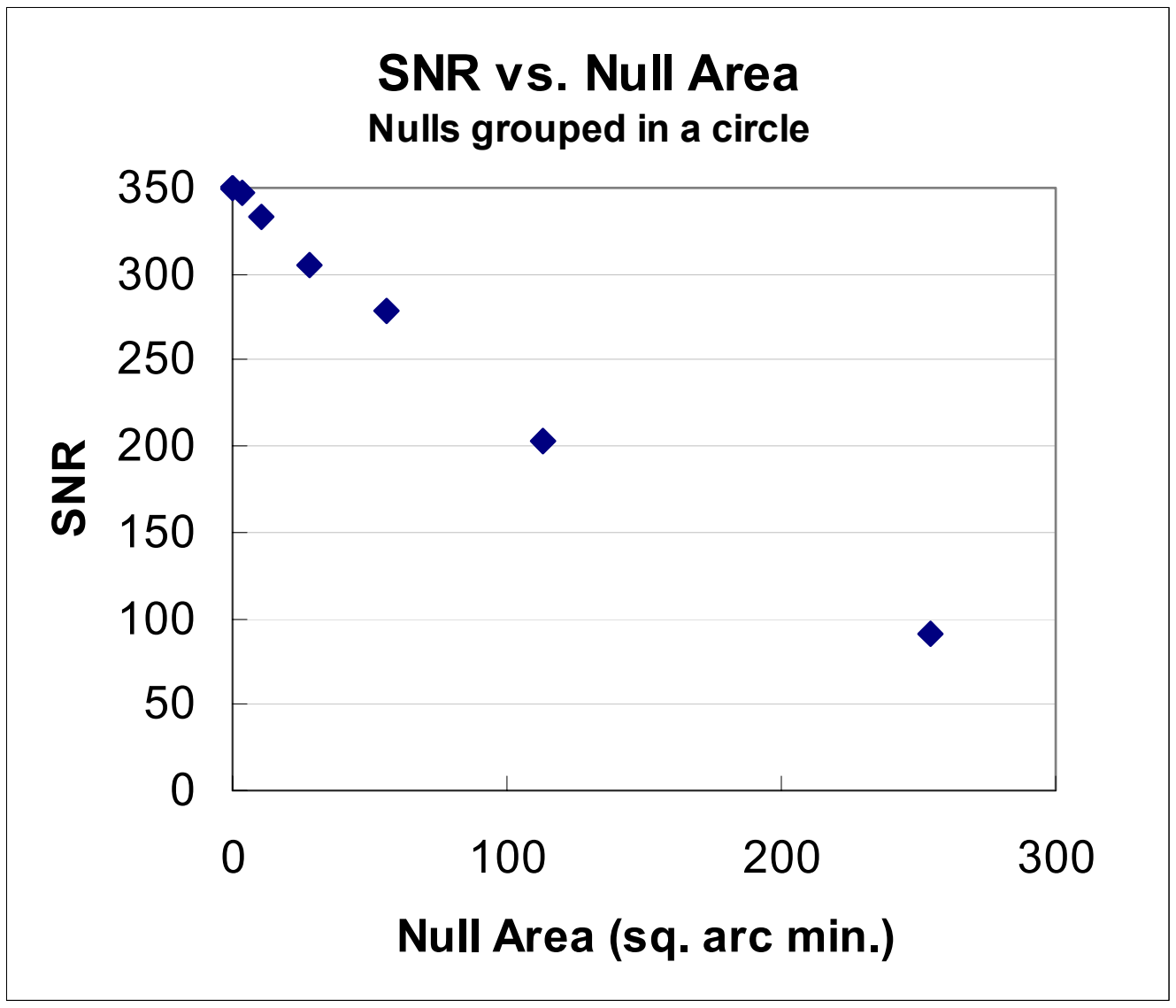

**Figure 7: The SNR on an unresolved point source for beams with nulls of various total area (that is, solid angle).**

[†] This is an over-estimate of the array diameter, as all the antennas fit easily into a 1 km diameter circle. Furthermore, the array is not symmetrical so $\Delta k$ is not really equal in all directions. Below we shall deduce a more accurate estimate of $\Delta k$ directly from the simulations.

Similar to the previous section, Figure 7 displays the SNR as a function of null area. The SNR drops almost linearly with area and we can achieve almost 100 sq. arc min. of null before dropping below SNR = 250.

It is interesting to examine the frequency dependence of the wide area null in Figure 6 (seen in Figure 8). As anticipated by the comparison in Figure 4, the wide-area of the present null is reflected in a wide BW if one focuses on a single point at 50° zenith angle. This demonstrates that WBW nulls are possible. However our results (not shown) indicate that the correspondence between angle and frequency breaks down for BW > 10 MHz. Next we explore better methods for WBW nulling.

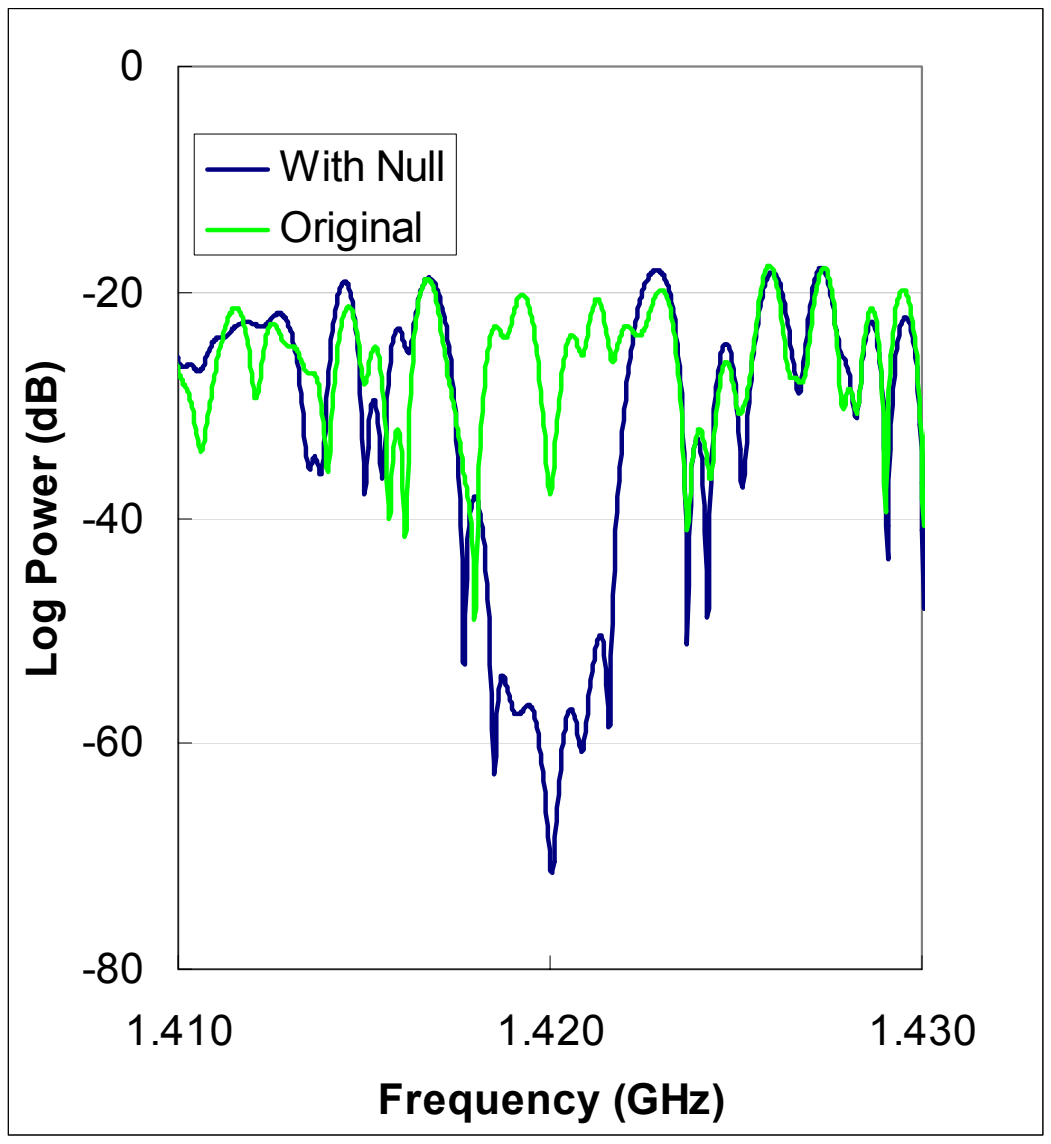

**Figure 8: A frequency scan through the same null as in Figure 7. This scan plots only the single point at null center. The wide area of this null translates to a WBW if one restricts focus to the beam center.**

# 6. Wide Bandwidth Nulls

By now, the generalization of the iterative technique to WBW nulls is straightforward. Equations 12 and 13 do not change at all; all that is required is to allow $|k_m| \neq k$ . That is, to create a null at a given direction and frequency, we simply calculate the gain at that direction and frequency. This value is used to correct the antenna weights via Equation 13. Just as before, iterations proceed until the desired level of nulling is reached.

## Example

Figure 9 shows an example of a WBW null calculated in this way. This is a single point null directed at 50° zenith angle in the $xz$ plane. A scan as a function of zenith angle and a scan as a function of frequency are shown. This beam has SNR = 324. As in the wide area case, we have chosen a null spacing of $\Delta k = \frac{2\pi}{L}$ , but this time along the frequency axis, corresponding to $\Delta f = 300$ kHz independent of tuning frequency. Figure 9 and other studies of ours shows that this choice is adequate to prevent the gain from rising above –50 dB between null points. Note too that this is six times larger than the frequency width of an isolated single point null. Again, isolated nulls seem to "occupy" a smaller region of $k$ than expected.

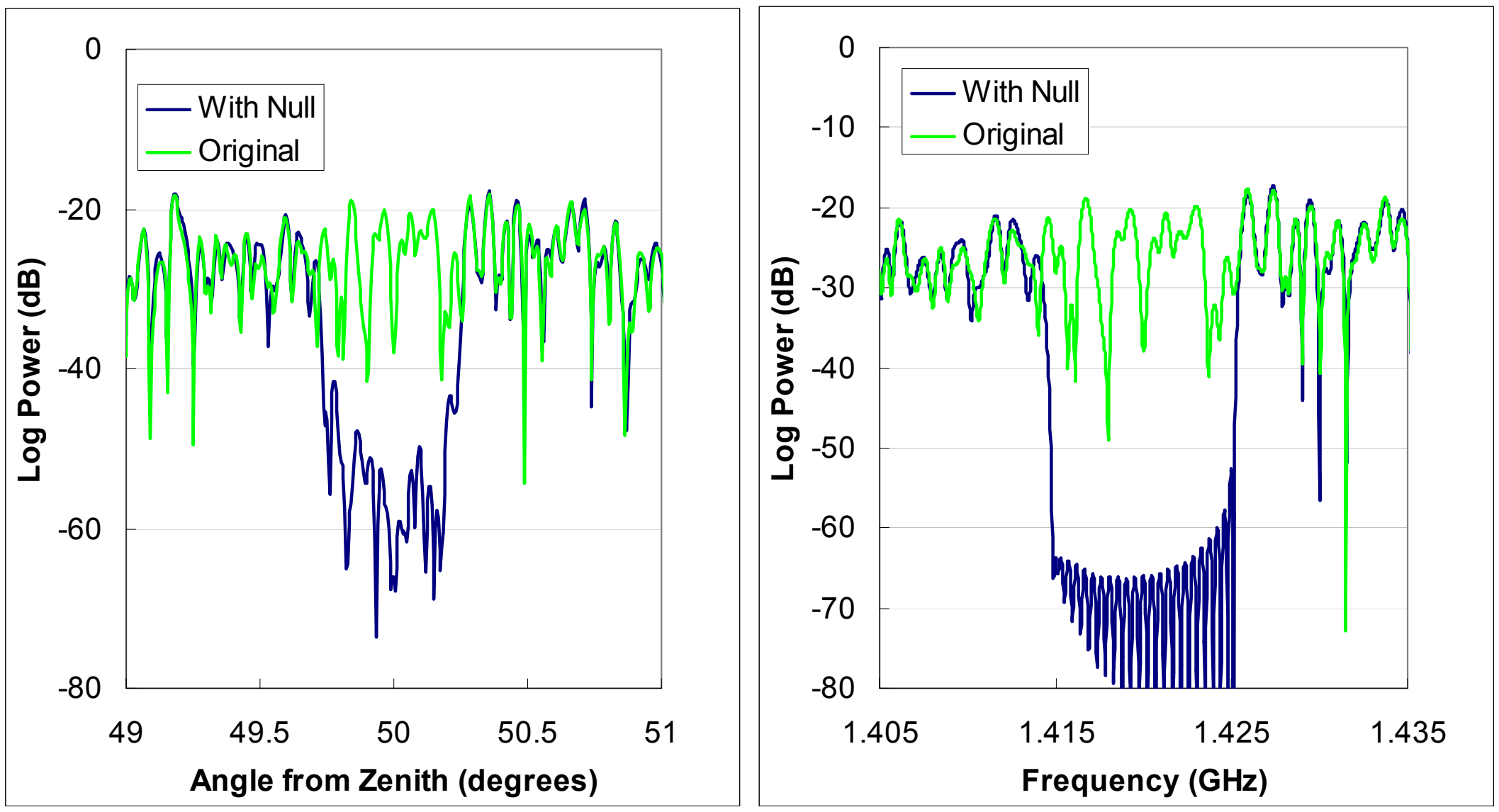

**Figure 9: Single point null over a WBW. The left panel displays an angle scan through the null at 1420 MHz, whereas the right panel displays the gain in the null direction as a function of frequency. Notice how the WBW at one angle translates approximately to a wide angle at one frequency.**

Of course, one is not restricted to creating nulls over a contiguous frequency range. In Figure 10 a null is placed at a comb of frequencies, with 1 MHz spacing over a range of 20 MHz.

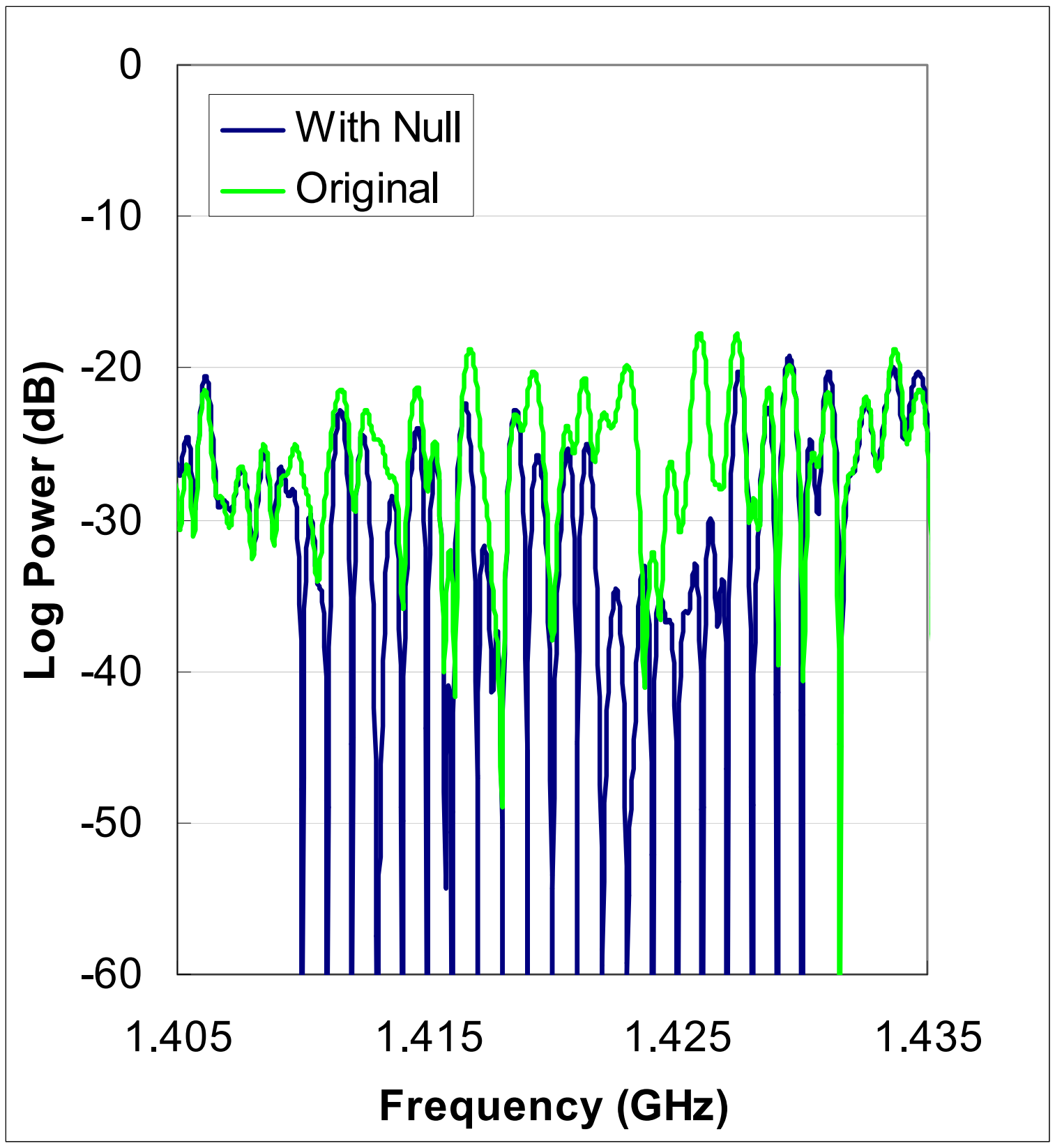

**Figure 10: Single point null placed over a comb of frequencies.**

Similar to section 5, we examine the SNR of a single point null as the BW is gradually increased in Figure 11. The SNR is decreases almost linearly with increasing BW.

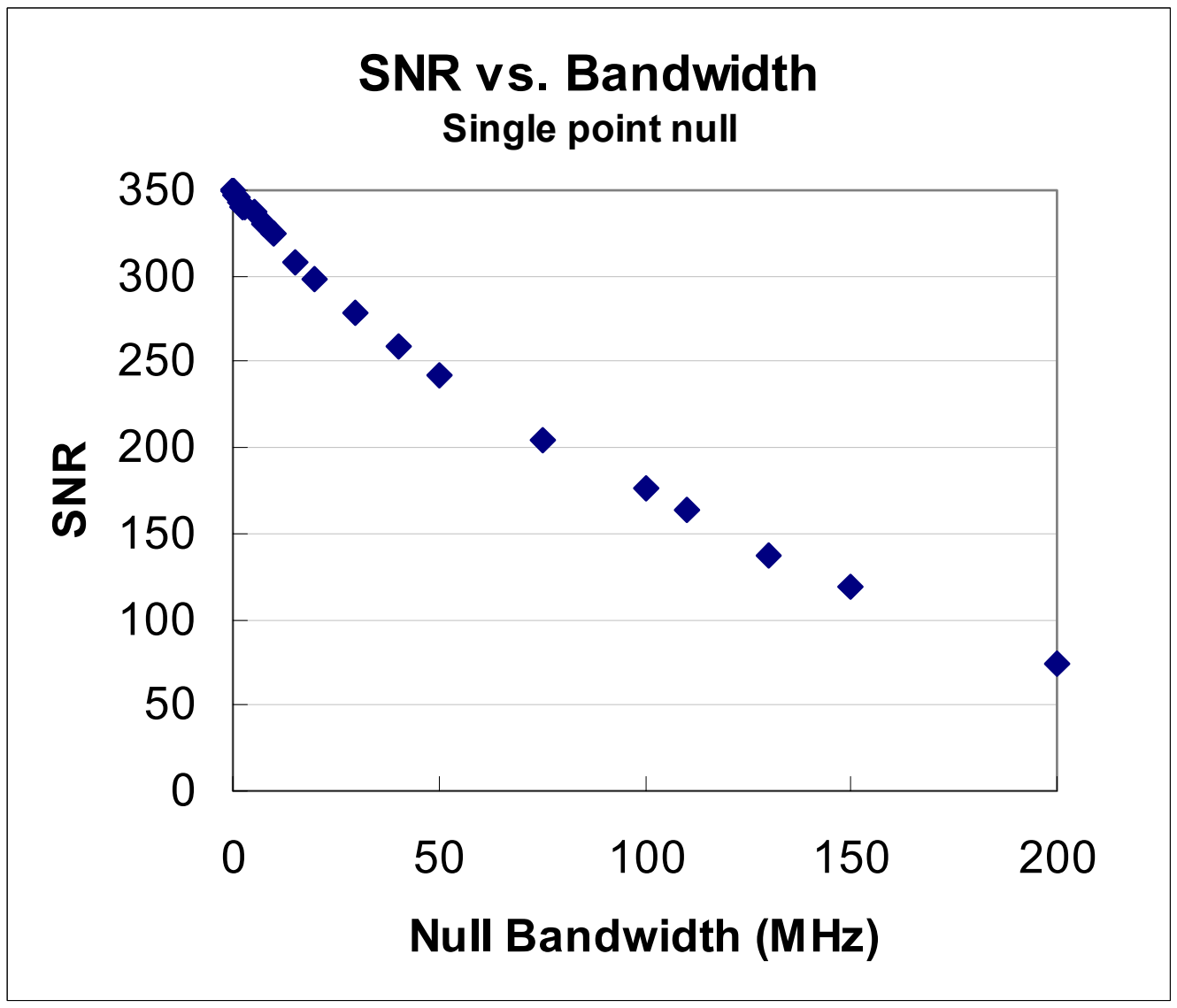

**Figure 11: Plot of SNR for a beam having a single point null placed with various BWs.**

# 7. Effects of Phase and Gain Miscalibration

In a real telescope such as the ATA, we will not know the exact gains or delays for the signals coming from each antenna. It turns out that this imprecise knowledge places serious limits on the depth of nulls that can be achieved in a beamformer. To illustrate, we considered a WBW point null similar Figure 9, but with a 3 MHz BW. This beam ("No Noise" in Figure 12) was initially created assuming perfect knowledge of the gain and phase for every antenna.

On the left hand side of Figure 12, the antenna coefficients were gradually perturbed with random Gaussian phase noise, i.e.

$$\text{Antenna Coefficient} \Rightarrow (\text{Perfect Coefficient}) \cdot e^{i\Delta\phi}$$

where $\Delta\phi$ is real and chosen from a normal distribution of values centered at zero and with standard deviation as shown in the figure. Similarly, the right hand side of Figure 12 shows the effects of amplitude (gain) noise by letting

$$\text{Antenna Coefficient} \Rightarrow (\text{Perfect Coefficient}) \cdot (1 + \Delta g)$$

where $\Delta g$ is real and chosen from a normal distribution of values centered at zero and with standard deviation as shown in the figure.

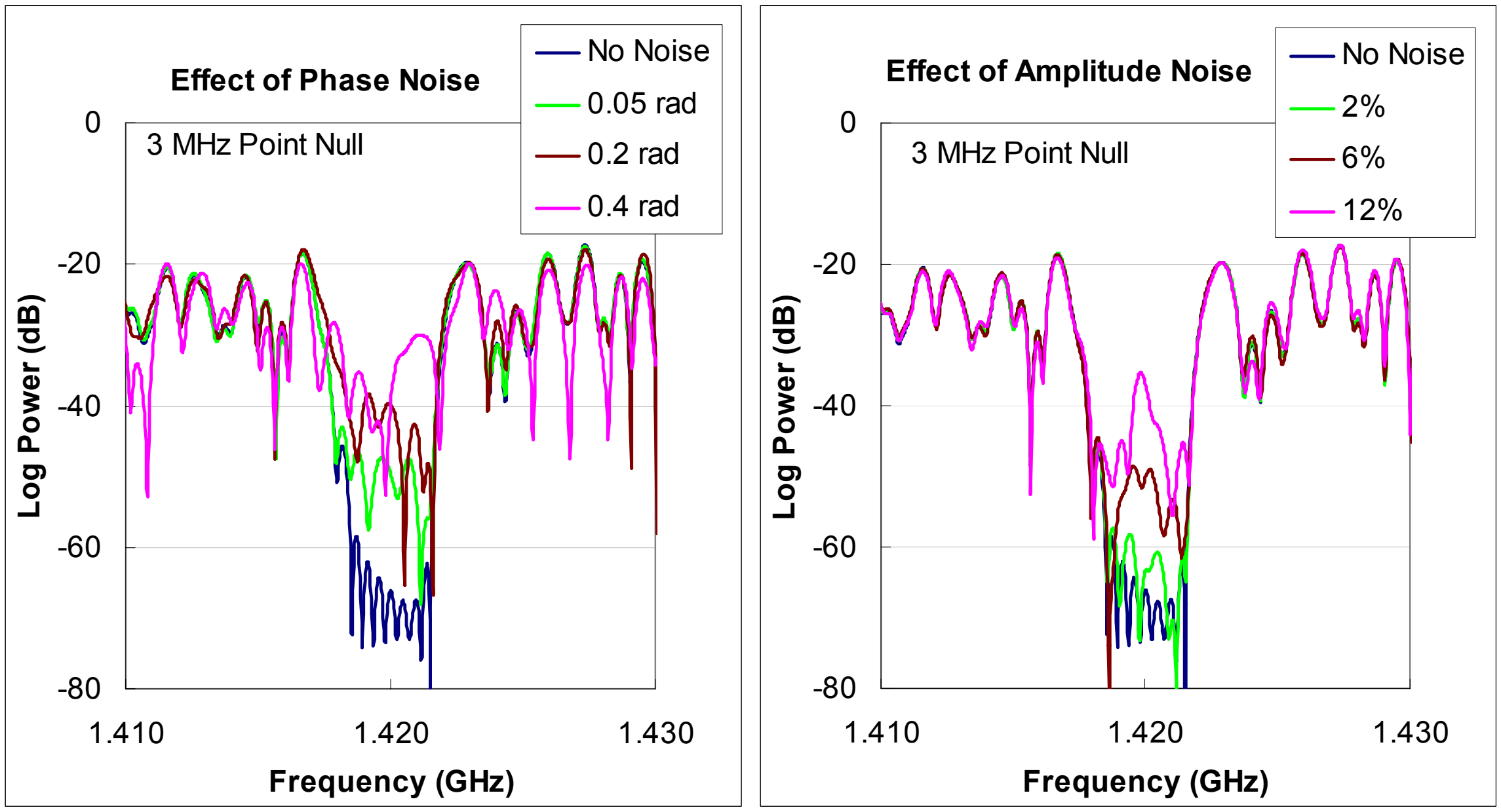

**Figure 12 Demonstrating the effects of phase noise or amplitude noise on the depth of a WBW point null. Initially (No Noise) the null has a depth < -60 dB below the main beam over 3 MHz BW. The antenna coefficients are perturbed with increasing amounts of phase or amplitude noise, and the resultant plots are shown.**

We observe that even the modest goal of achieving 40 dB of suppression relative to the main beam is attainable only with very accurate knowledge of the signal phases (better than 0.2 radians). Comparatively, the required gain accuracy (better than 10%) appears more easily achieved. Of course, gain and phase errors are cumulative, so a balance must be struck between them when both are present (as at the real ATA). We have undertaken similar studies of wide area and isolated point nulls, both of which lead to similar conclusions.

## 8. Shaping the beam

In the previous section we showed that null patterns can be created by simply "drawing" on the calculated gain pattern of the array. In this section we follow a similar procedure except that instead of drawing the nulls, we draw in the desired shape of the beam itself. It is of interest to see how much control over the beam shape we can exert without losing too much sensitivity.

As an example, we consider the "ring beam," which is a construction that might have utility in pulsar and / or SETI studies. This beam pattern has a deep null in the pointing direction surrounded by a ring of high gain. This type of beam might be used in on / off studies, where one beam is centered on a star while a ring beam simultaneously measures the signal in the vicinity of the star without including any signal from the star itself.

The method for generating the ring beam is straightforward. Begin by drawing the desired beam pattern as in Figure 13, left. Inverse fourier transform this beam pattern to obtain the "theoretical array" similar to that in Figure 1. Now project out the "real" antenna coefficients from this theoretical array. This is already the good approximation to the desired beam shape. In the case of the ring beam shown here, we used these coefficients as the starting point for an iterative cycle to reduce the gain to zero right at the center of the ring (since there is only one null, only one iteration is required).

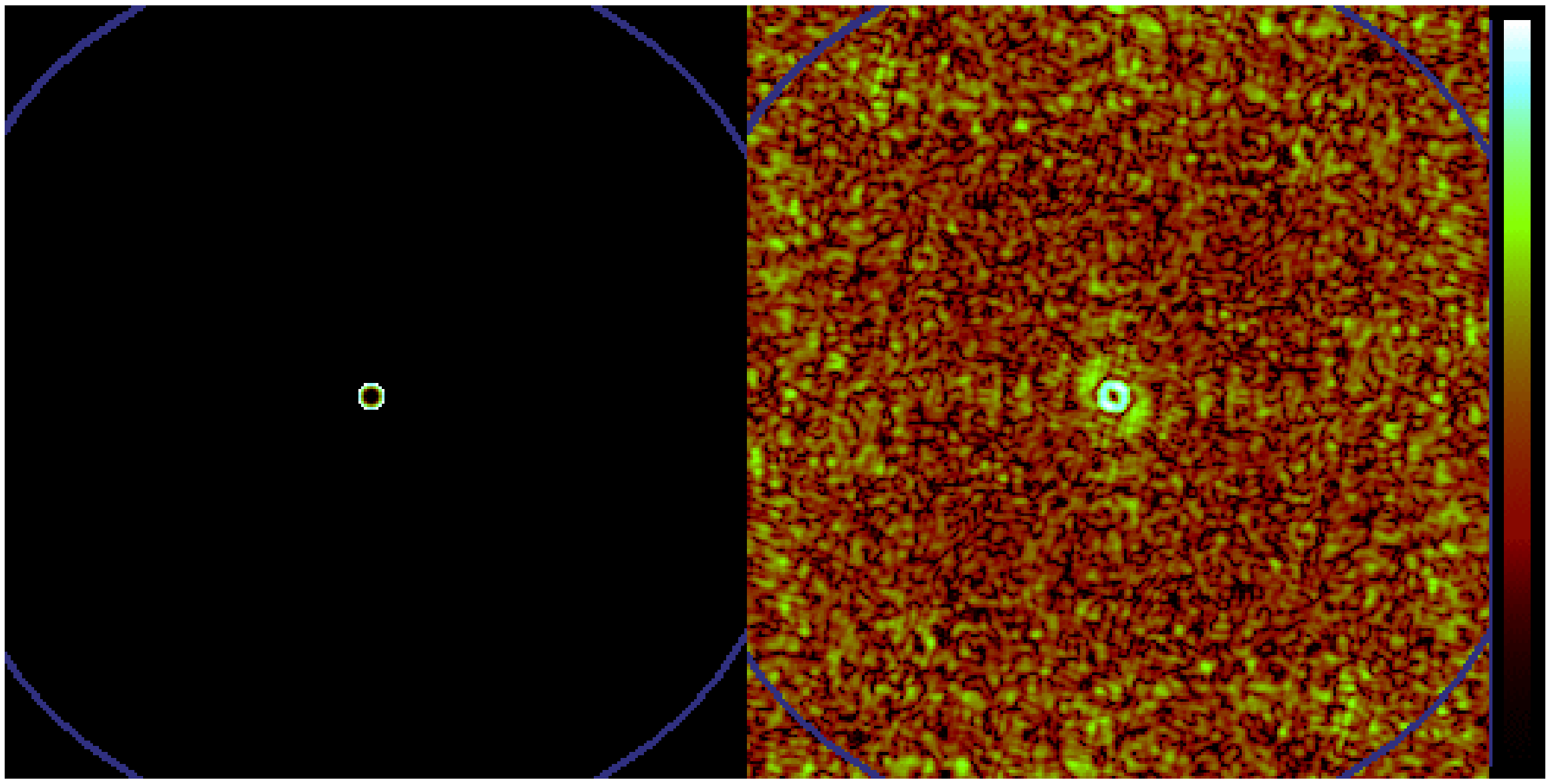

**Figure 13: Left: Desired beam pattern for ring beam. Right: Actual ring beam as calculated.**

We compare linescans through an ordinary beam centered at zenith with a ring beam centered in the same direction (Figure 14). The ring beam has a deep null (< -100 dB) exactly at beam center. These beams are normalized such that they have the same integrated power over their central regions (out to first minima excluding ring beam minimum at zero angle). The ring beam has noticeably higher sidelobes, which is an indicator of loss in sensitivity. Nevertheless, the ring beam can be a useful tool for the radio astronomer. Apart from the ring beam, other beam shapes come to mind such as disks for planets, or generic shapes for galaxies or nebulae.

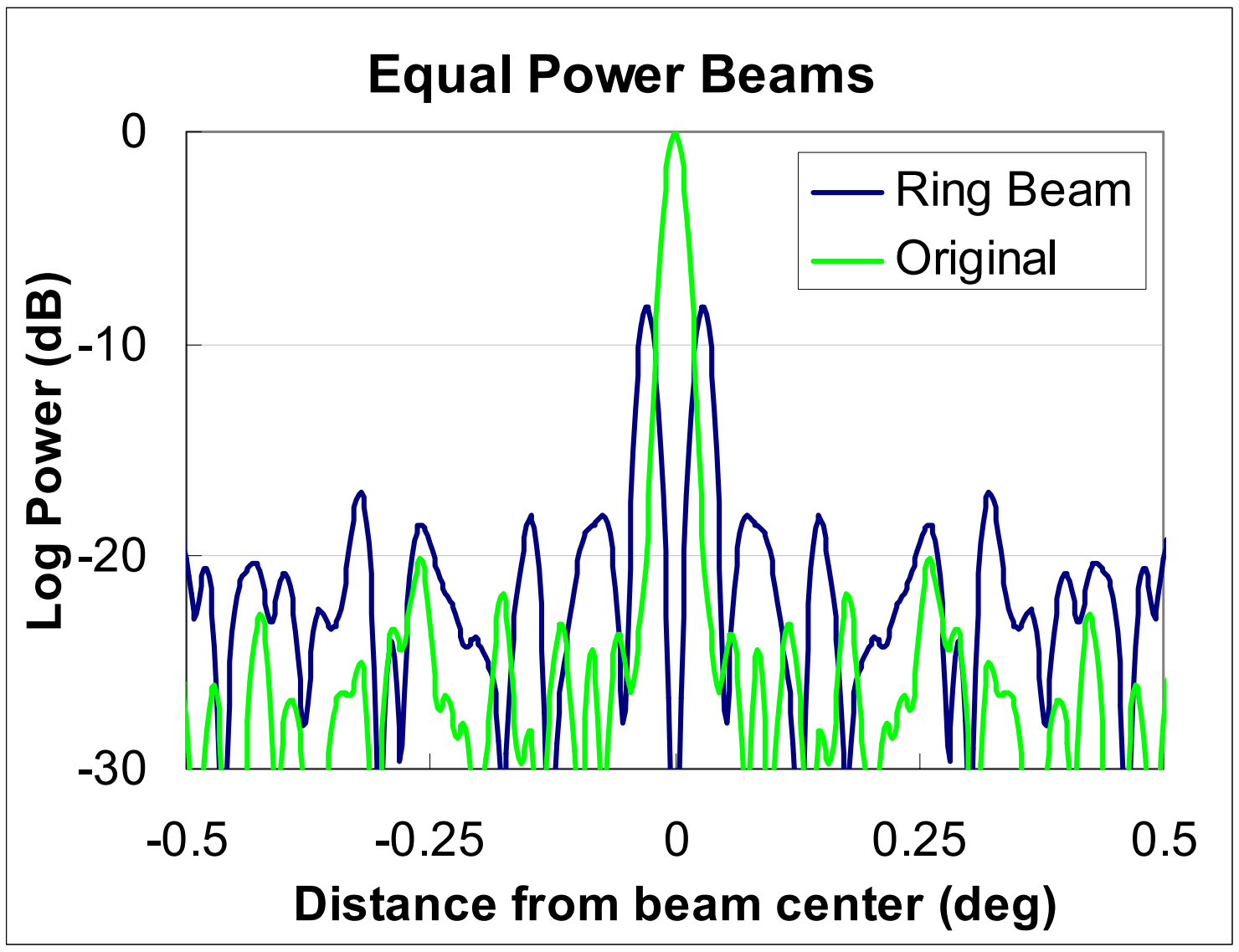

**Figure 14: A comparison of an ordinary beam centered on an object with a ring beam centered on the same object.**

For all shaped beams there are an infinite number of solutions obtainable from the method described above. This is because the "origin" of the antenna pattern is arbitrary. That is, we can place the origin in antenna space to be coincident with Antenna 1, Antenna 350, or anywhere. But the antenna origin has a strong impact on the values of the antenna coefficients as follows.

First consider the inverse fourier transform of the desired ring beam. This is a pattern having high amplitude at the center and surrounded by concentric rings of amplitude that decays with increasing radius. Similarly the phase has azimuthal symmetry and varies smoothly with radius. Now the array of antenna positions is a "cookie cutter" which we place over this pattern. We can place the cookie cutter anywhere on this pattern and punch out different coefficients. We do not know of a way to characterize the SNR of a ring beam such as in Figure 14, but we are certain from empirical results that some choices of array origin provide better results than others. Future work may consider this problem more closely.

# 9. Discussion

## Control of Coefficient Amplitude

In order to achieve wide area / WBW nulls, it is necessary to exert full control over the amplitude of the antenna coefficient, as seen in Figure 15. Here we display the amplitude of the antenna coefficient as a function of distance from the (arbitrary) array center for several null configurations. The amplitudes have been normalized so that the largest among the group is unity. It is seen that for small nulls with small BW, the amplitudes are all close to unity (similar to the results of Ref. 4). However, as the null diameter increases, the amplitude variation increases and many antenna coefficients have amplitude close to zero.

We do not claim that these results are new or even unexpected. However, this is an important point for the ATA, since one proposed design of the signal processing hardware would allow high-speed control of the antenna amplitudes only over a range of ± 12% of unity[5]. Thus if we are to have the opportunity to test ideas for wide-area nulling at the ATA, it is critical to have full high-speed control over signal amplitude. The same statement is true for controlling the shape of the forward gain as in Section 8.

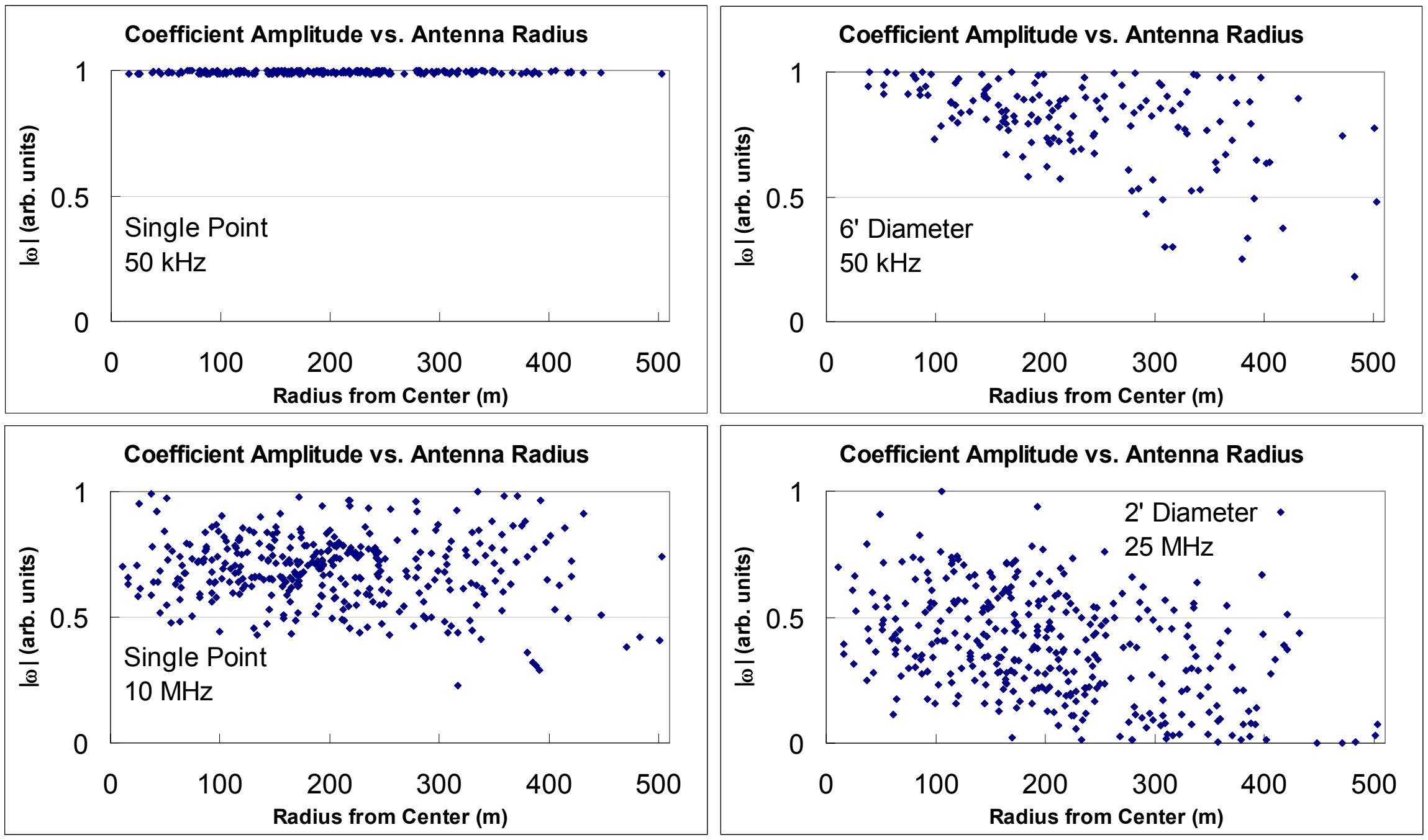

**Figure 15: Coefficient amplitude as a function of radius from the array center. Each panel corresponds to a null diameter and BW as indicated.**

## SNR versus $k$-space Volume

One question that arises is how much nulling can be tolerated without compromising the astronomical measurement. For a selection of nulls widely spaced in angle (or frequency), this question is pretty much answered with Figure 5. But if the RFI position is not precisely known and / or one wishes to null over a range of frequencies, then the answer more complicated.

The three-dimensional space spanned by all possible values of $\vec{k}$ as defined here ($k$-space) leads to a natural resolution. In this space, radius from the origin corresponds to frequency, with DC at the origin. The beam pattern over the entire sky at a narrow BW resides in a spherical shell of $k$-space.

Furthermore, each point null might be regarded as occupying the volume element $\Delta k^3 = \left(\frac{2\pi}{L}\right)^3$ centered on the tip of the vector $\vec{k}_m$. Perusing this thought, we calculate the total volume $\Delta V_k$ occupied by a wide area / WBW null as follows. The wide area nulls considered here represent disk shaped regions confined to spherical shells in $k$-space. These disks have (frequency) thickness $\Delta k$. A single point, WBW null follows a radial line segment with length corresponding to the BW, and having areal cross section $(\Delta k)^2$.

With these presumptions we may plot the SNR from Figure 5, Figure 7, and Figure 11 on a single figure, versus $\Delta V_k$ (Figure 16). Here we see the wide area and WBW SNR curves collapse to the same curve with the introduction of no free parameters.

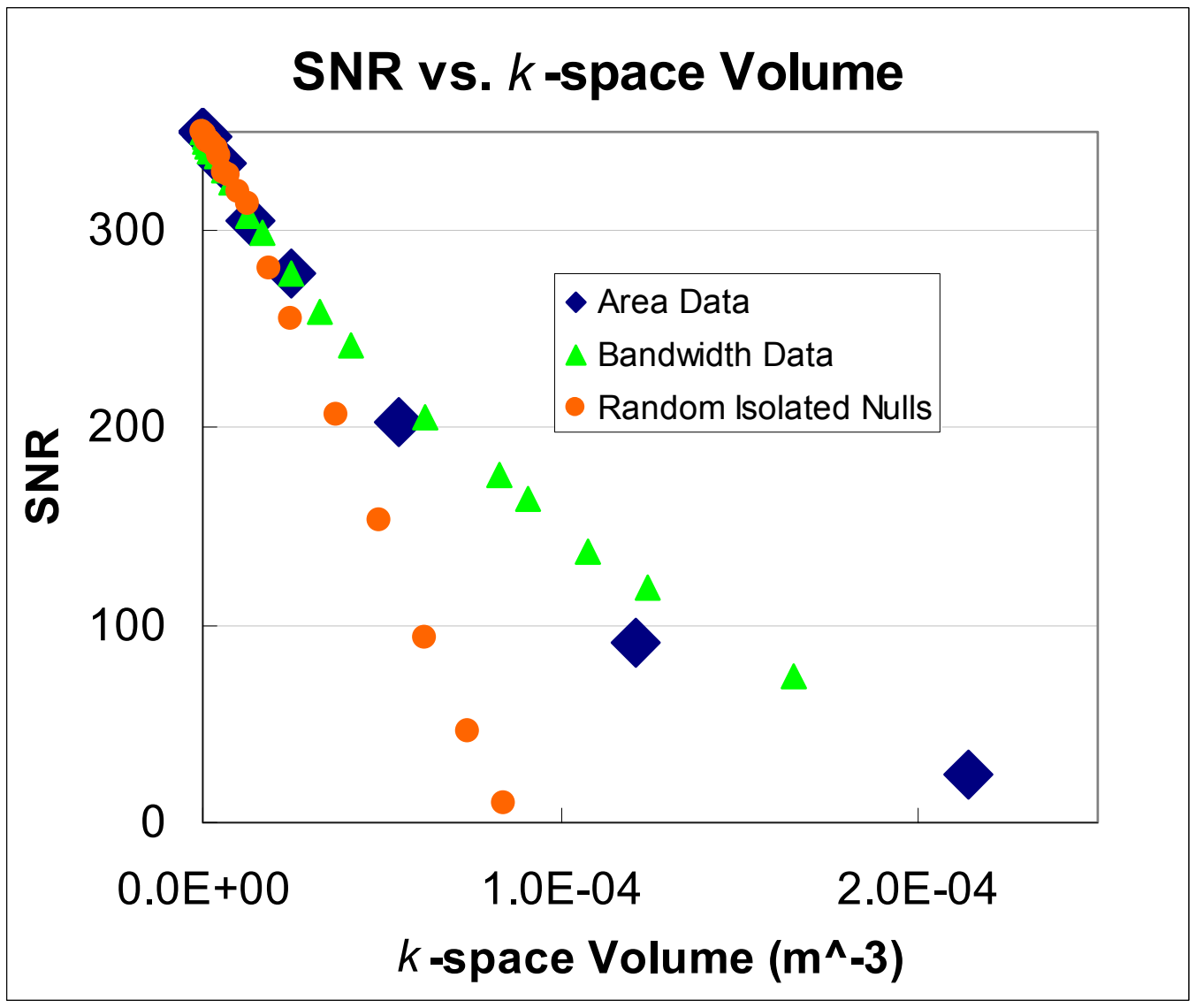

**Figure 16: SNR from figures 5, 7, and 11 plotted versus $\Delta V_k$, the volume occupied by these nulls in $k$-space. Without introducing any fitting parameters, the wide area and WBW SNR curves collapse to a line.**

The SNR from the randomly placed nulls does not agree with the other curves. We explain this mismatch by our approximation that the characteristic array spatial dimension $L$ = 1000 m. Actually, the array is smaller. It is difficult to estimate $L$ directly because the array has an irregular shape so we deduce its value by demanding that all three curves come into alignment. Good alignment is found for $L$ = 750 m (Figure 17), which is reasonable considering that the longest baseline is ~ 1000 m and that a square array with diagonal 1000 m has $L$ = 707 m.

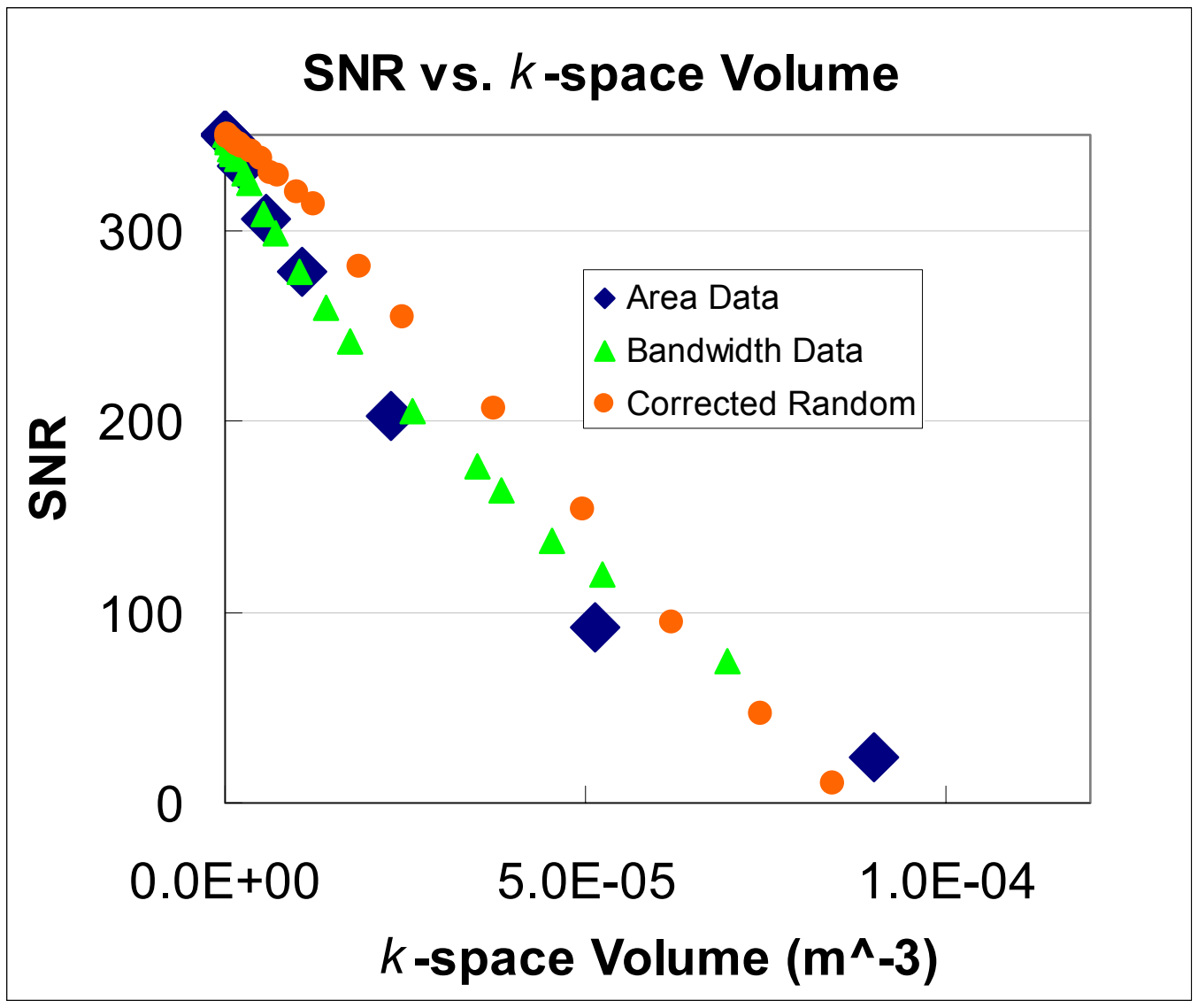

**Figure 17: Using the array characteristic length as a fitting parameter, the SNR curves of Figure 16 are brought into alignment.**

The evident linear relationship between SNR and $\Delta V_k$ suggests that in general, SNR can be predicted via this formula:

$$\text{SNR} = N - \xi\, \Delta V_k = N - \xi\, M\, \Delta k^3 \quad , \tag{16}$$

where $M$ is the number of independent null points. To deduce $\xi$, we note that in the isolated null case we have SNR = 0 when $M = N$, leading to

$$\text{SNR} = N - M\ . \tag{17}$$

Eq. 17 indicates that in terms of SNR, each independent narrowband null effectively removes one antenna from the array.

As in the discussion relating to Figure 15, Eq. 17 is not new and can be derived directly from the linear algebraic interpretation of beam forming.[6] Nevertheless this point is often forgotten in the excitement surrounding null steering. Even a well-understood interferer like a GPS satellite requires ~20 individual point nulls to be effectively removed, because its transmission bandwidth exceeds $\Delta k$ (300 MHz) by the same factor. Hence the null steering method can be a relatively expensive RFI mitigation strategy.

By comparison, another RFI strategy that has been given much attention is to sacrifice one antenna to point directly at the interferer and measure its signal very well. Then, by correlating the interference signal with the output of the beamformer (over a range of time lags), one estimates the interference signal in the beamformer output. Finally, the interference signal is subtracted out. The latter approach sacrifices one antenna but has the potential of removing RFI over a wide range of frequencies. By comparison, the null steering method requires the sacrifice of many antennas if one wishes to remove RFI over a broad frequency range.

# 10. Conclusions

Here we have presented a novel approach for deducing null steering coefficients that has led to a few useful insights, especially that of a prescription for wide bandwidth null steering. We have explored the limits of size and bandwidth that can be accomplished using beam formers in the proposed design of the ATA. One important implication for ATA design is the finding that full high-speed amplitude control over antenna coefficient amplitude is necessary for successful application of null steering and / or beam shaping methods.

## Acknowledgements

The author gratefully acknowledges G. C. Bower for many useful discussions of RFI mitigation techniques, S. W. Ellingson for helpful criticisms and encouragement, M. Wright for conversations regarding beam forming, M. Davis for critical review of an early version of this manuscript, and R. F. Ackermann for several useful discussions. Without the knowledge and creative input of all these scientists, this work would not have been possible. The author is also grateful to the SETI Institute for support of the present work.

# Appendix: Some whimsical null patterns

Here we reproduce a series of nulling patterns that were created to demonstrate the flexibility of null placement that is available in the proposed ATA design. All of these beams have relatively low SNR because most of the degrees of freedom have been absorbed by the nulls.

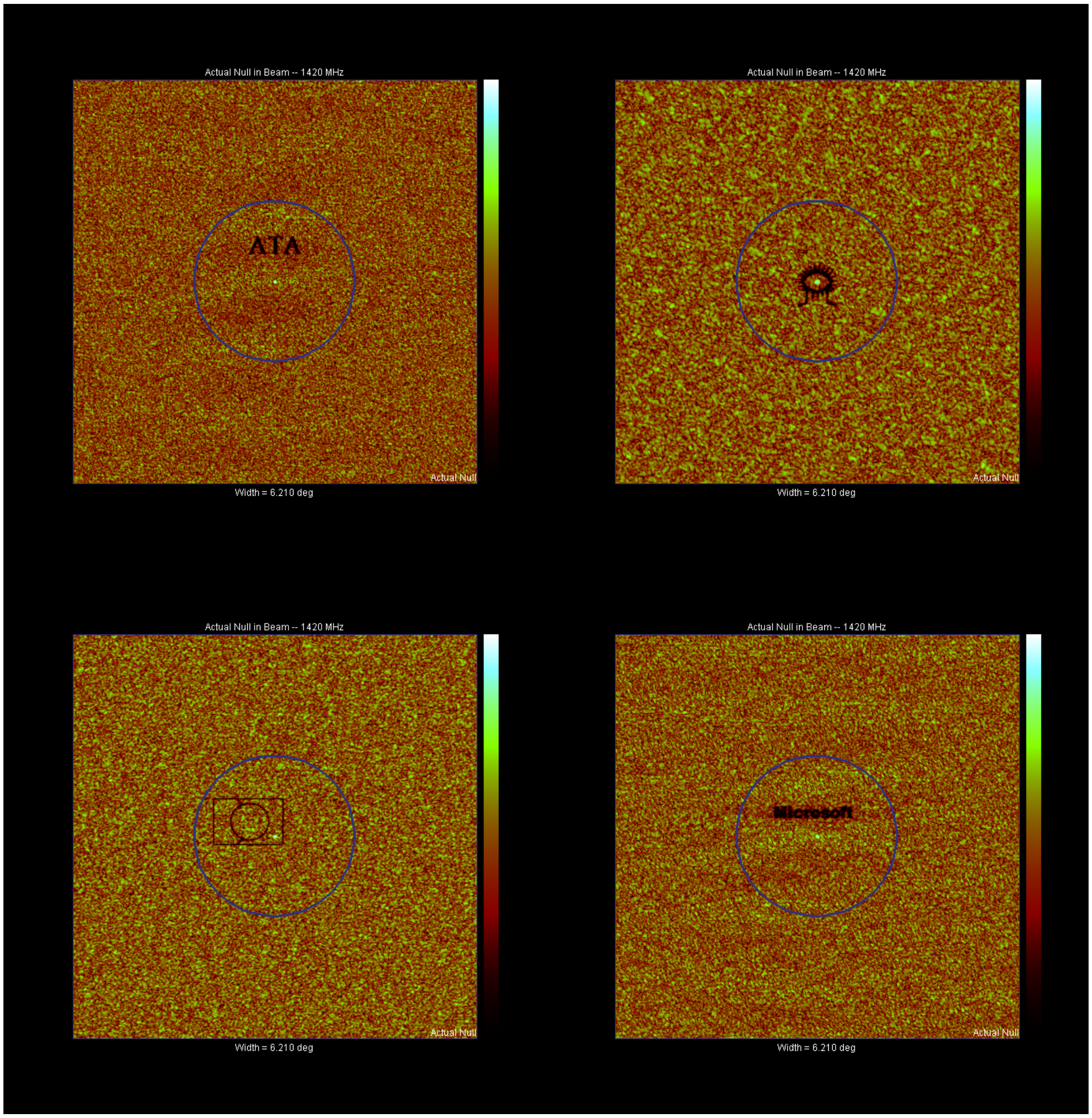

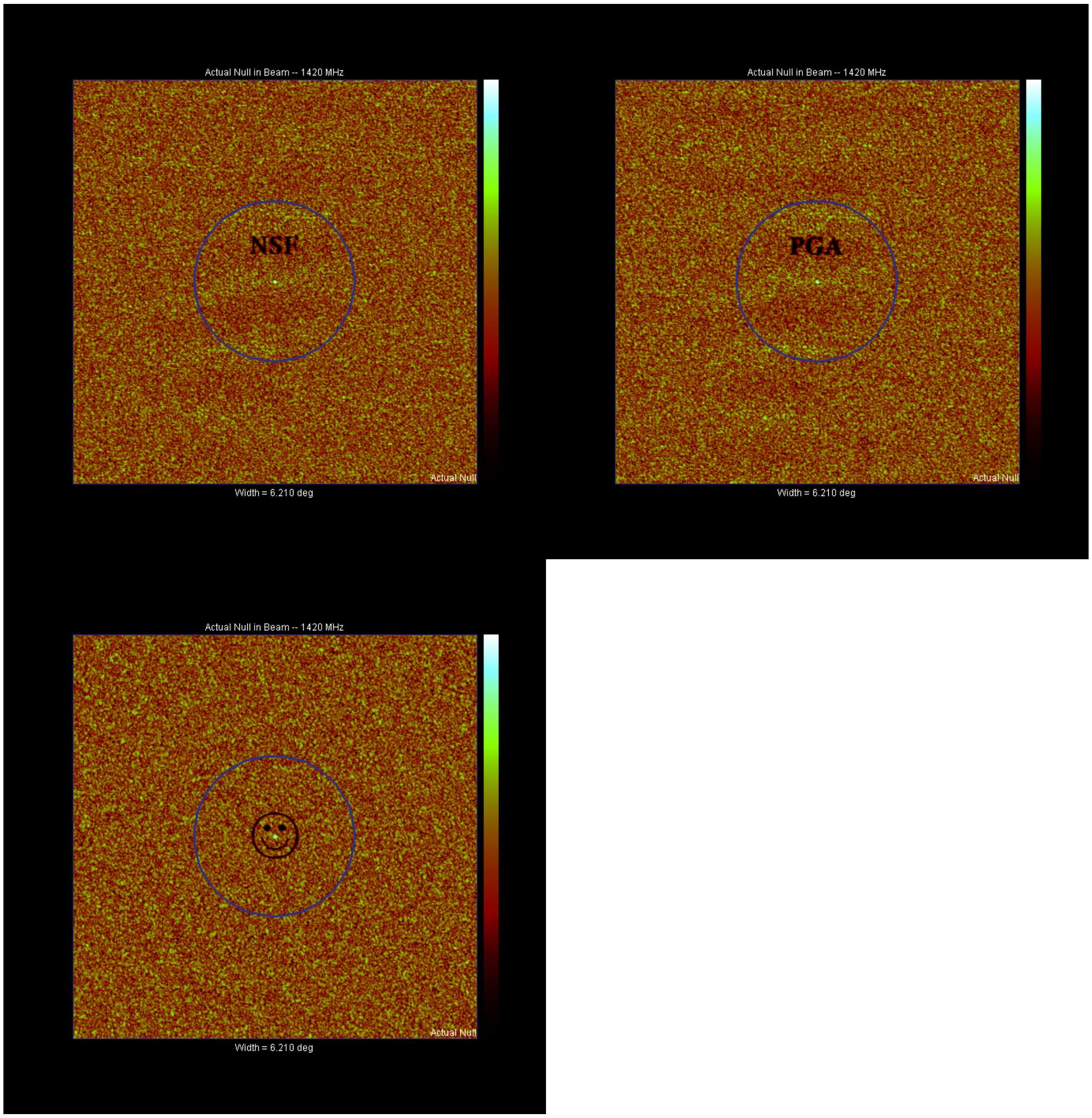